\documentclass[prb,twocolumn,showpacs,showkeys]{revtex4}
\usepackage{graphicx}
\usepackage{dcolumn}
\usepackage{amsmath}
\usepackage{amssymb}
\usepackage{bm}
\begin{document}
\newcommand{\be}{\begin{equation}}
\newcommand{\ee}{\end{equation}}
\newcommand{\vk}{\mathbf{k}}
\newcommand{\vq}{\mathbf{q}}
\newcommand{\vp}{\mathbf{p}}
\newcommand{\vx}{\mathbf{x}}
\newcommand{\rsa}{r_{\text{sa}}}
\newcommand{\lsa}{\lambda_{\text{sa}}}
\newcommand{\half}{\frac{1}{2}}
\newcommand{\la}{\langle}
\newcommand{\ra}{\rangle}
\newcommand{\caH}{\mathcal{H}}
\title{Interplay between local electron correlation and Janh-Teller electron-phonon interaction}
\author{Choong H. Kim and Hyun C. Lee}
\email[Corresponding author.Electronic Address:~]{ hyunlee@sogang.ac.kr}
\affiliation{Department of Physics and Basic Science Research Institute, 
Sogang University,Seoul, 121-742, Korea}
\date{\today}
\begin{abstract}
The infinite-$U$ Anderson model coupled to  a Jahn-Teller phonon is studied using the slave boson method on the basis of 
 the large degeneracy expansion ($1/N$) scheme. The model Hamiltonian acts on the orbital degrees of freedom. 
The main focus is on the interplay between strong local electron
correlation and weak Jahn-Teller electron-phonon interaction. 
The Kondo temperature is found to decrease by Jahn-Teller interaction. The influence of the Jahn-Teller interaction on dynamical
correlation functions is very significant  in sharp contrast with the case of  the Holstein-type phonon which
couples to charge degrees of freedom.
\end{abstract}
\pacs{71.27.+a, 71.10.-w, 63.20.Kr, 71.38.-k}
\keywords{strong electron correlation,Jahn-Teller interaction,Kondo physics}
\maketitle
The orbital degrees of freedom of electrons in a solid play a prominent role in 
modern material science.\cite{orbital1}
In particular, it turns out that  electric current can be controlled by manipulating the orbital states of electrons. 
Spurred by this possibility, a new  research area  called \textit{orbitronics} is emerging.\cite{orbital2}
The orbital physics is also relevant to the quantum transport through orbitally active atoms (or defects)
 in nanostructures.\cite{park2002,clough}

The dynamics of the orbital degrees of freedom is determined by the Coulomb repulsion between orbitally active electrons and 
the coupling to the vibrations of ions surrounding the electrons. 
A well-known example where the orbital physics manifests itself
 is the transition-metal oxides.\cite{orbital1} Electrons of these materials occupy $d$ orbitals, 
and the Coulomb repulsion between these electrons is \textit{very strong}.
At the same time the electrons in $d$ orbitals are coupled to the vibrations of surrounding oxygen atoms via the so-called
Jahn-Teller (JT) coupling.\cite{JT} The JT coupling stems from the anisotropy of $d$ orbital wave functions.
The \textit{orbital pseudospin} operator to which JT phonon couples is defined by
\be
T_{ \alpha}= \sum_{\sigma} \sum_{a,b} c^\dag_{ \sigma a} \tau^\alpha_{ab} c_{ \sigma b},
\ee
where $\sigma$ is a real spin index and $a,b$ are orbital indices.
For the case of $e_g$ orbitals of $3d$ electrons, the orbital indices $a,b$ designate two orbitals
$d_{x^2-y^2}$ and $d_{3z^2 -r^2}$, and the matrices $\tau^\alpha$ are  the  $2\times 2$ Pauli matrices.

The progress of  theoretical understanding of orbital physics has been slow in spite of its  prime importance.
The reason is that the physical phenomena of orbital physics are usually associated with \textit{strong} electron correlation
and \textit{simultaneously} with  strong coupling to phonons of various types. This makes theoretical analysis exceedingly difficult.
Despite these difficulties, the phase diagram of manganese oxides could be qualitatively understood 
by the mean-field approximation study on the generalized 
Hubbard model,\cite{meanfield} but the understanding of \textit{dynamical} properties 
in low energy strong coupling regime is beyond the reach of such mean field approximation. In passing we mention
 that for \textit{weakly} interacting electron systems  there exists a very successful Migdal-Eliashberg theory.\cite{ME}

To understand orbital physics in the presence of strong correlation  we need  the approximation schemes which preserve 
the essential features of strong correlations. 
One of such approximation schemes is the dynamical mean-field theory (DMFT).\cite{dmft1,dmft2} In DMFT  
a lattice Hamiltonian is mapped (or approximated)
to a certain  \textit{quantum impurity} Hamiltonian. 
In this brief report we study an Anderson impurity Hamiltonian interacting with local JT phonons (AJT model)
from the perspective of the orbital physics within the DMFT scheme or the transport phenomena in nanostructures.

To simplify our problem further \textit{we will assume that real spins are polarized}, namely, the material under consideration
is in a ferromagnetic state.
This simplification is not unrealistic, since a large portion of phase diagrams of transition-metal oxides and multiferroics
are ferromagnetic. A typical example  is the low temperature phase of
$\text{La}_{0.7}\text{Sr}_{0.3} \text{Mn} \text{O}_3$ manganese compound with $T_c \sim 370~\text{K}$ 
which shows the colossal magnetoresistance.\cite{urushibara}
An additional advantage of studying the ferromagnetic state is that 
we can  focus on the orbital dynamics itself  not being mixed with spin dynamics.

The physical properties of the AJT model become clearer when they are compared with those of 
Anderson-Holstein (AH) model.\cite{holstein,hewson,hewson2,jpc,ours,hyun2004}
The dynamical degrees of freedom of the AH model is the \textit{real spin}. The orbital degrees of freedom is not present in AH model.
Most important difference between two models is that the Holstein phonon couples to the \textit{charge} degrees of freedom of impurity
electron while JT phonon couples to \textit{orbital} degrees of freedom.
It is well known that the nonperturbative Kondo singlet ground state of the Anderson model (in local moment regime) is due to the quantum
fluctuations in spin  
(or orbital pseudospin in the context of AJT model) channel. In this regard one can expect that the influence of Holstein phonon
on the Kondo ground state would be  \textit{small} unless the electron-phonon coupling is extremely large, 
and this expectation has been confirmed.\cite{hewson,hewson2,jpc,ours,hyun2004}
On the other hand the influence of JT phonon is expected to be substantial since it directly couples to the channel responsible for
Kondo (orbital pseudospin) singlet ground state. The results of our study confirm this expectation.
 
The AJT Hamiltonian consists of the following three parts:
\be
\label{hamil1}
\caH_{\text{AJT}} = \caH_{\text{el}} + \caH_{\text{ph}}+\caH_{\text{JT}}.
\ee
$\caH_{\text{el}}$ and $\caH_{\text{ph}}$ are the Hamiltonians for the electron and phonon parts, 
respectively. $\caH_{\text{JT}}$ is the
Hamiltonian describing the interaction between the impurity electron and the JT phonon,
\begin{align}
\label{hamil2}
\caH_{\text{el}} &= \sum_{k,a}(\epsilon_k-\mu_c)~c^\dag_{k a} c_{ka}
+ \epsilon_f \sum_a f^\dag_a f_a + U f^\dag_\uparrow f_\uparrow  f^\dag_\downarrow f_\downarrow \nonumber\\
&+ \sum_{k a} \frac{1}{\sqrt{N_{{\rm lat}}}} ( V_k f^\dag_a c_{k a}+ V_k^*c^\dag_{k a} f_a),
\end{align}
\begin{equation}
\label{hamil3}
\caH_{\text{ph}} = \frac{1}{2} \left(\frac{P^2}M +M\Omega^2Q^2\right),
\end{equation}
\be
\label{hamil4}
\caH_{\text{JT}} = g_0 Q T_z,\quad T_z = \half
(f^\dag_{\uparrow} f_{\uparrow}- f^\dag_{\downarrow} f_\downarrow).
\ee
The  indices $a=\uparrow,\downarrow$ denotes the orbital degrees of freedom
as $\uparrow=d_{x^2-y^2}$ and $\downarrow=d_{3z^2-r^2}$, and the real spins are assumed to be perfectly polarized.
$f_a$ is the local impurity electron operator,and $c_{k a}$ is the conduction electron operator.
$V_k$ is the hybridization matrix element, $\mu_c$ is the chemical potential for the conduction electrons,
$\epsilon_f$ is the energy of the impurity level, and $U$ is the local Coulomb repulsion at the impurity site.
$T_z$ is the $z$ component of orbital pseudospin.
$Q$ is the local Jahn-Teller phonon coordinate. $P$ is the conjugate momentum of $Q$ satisfying
$[Q,P]=i \hbar$, and
$M$ is ion mass.
The JT phonon is assumed to be an Einstein phonon with frequency $\Omega$.
$N_{\text{lat}}$ is the number of lattice sites for the conduction electrons.
$g_0$ is the electron-phonon coupling constant which is assumed to be \textit{weak}.

>From Eq. (\ref{hamil4}) one can notice that the JT phonon plays a role of the 
fluctuating magnetic field acting on orbital pseudospin. The coupling to the magnetic field
evidently hinders the formation of the Kondo singlet ground state, thus  the Kondo temperature
at which the singlet begins to form is anticipated to decrease. Our results will show that this anticipation is indeed correct.
The electron-phonon interaction of the AH model is $
\caH_{\text{Hol}}=g_0 Q \rho,$ where $\rho = \sum_a \, f^\dag_a f_a$ is the impurity \textit{charge} density, and the index $a$
is the \textit{real spin} index.

The essential features of the Kondo physics of the Anderson model are kept intact in the limit of 
infinite local Coulomb repulsion $U$.\cite{thebook} In this limit the doubly occupied impurity state is not allowed 
in the physical Hilbert space. The elimination of doubly occupied state provides a considerable
formal simplification in the framework of the slave boson method.\cite{coleman,read,read2} Furthermore it
 possesses the great
advantage that it can describe the nonperturbative Kondo singlet ground state at the mean-field level. 
We  study the model Hamiltonian $\caH_{\text{AJT}}$ in the limit of infinite Coulomb repulsion, 
employing  Coleman's slave boson method in the scheme of $1/N$ expansion\cite{coleman,SG84}
($N=2$ for the original $\caH_{\text{AJT}}$).
In this limit the impurity electron operator $f_a$ can be expressed as
\be
f_a^\dag = s_a^\dag b,
\ee
with a constraint $b^\dag b + \sum_a s_a^\dag s_a = 1$.
$b$ is the slave boson operator, and $s_a$ is a constrained fermion operator.
 The above constraint is implemented by a complex Lagrange multiplier
 $\lambda = i \Omega_0 + \lsa$.
Next we take the  Read-Newns gauge in the scheme of $1/N$ expansion
in the form $s_a=z_a e^{i \theta}, b=(\sqrt{N/2}) r e^{i \theta}$. The conduction electron and 
JT phonon can be integrated out exactly, and then
we are led to the following (imaginary time) effective action:
\begin{align}
\label{action:rn1}
S &=\int_0^\beta d \tau \, \sum_m z_a^{\dag} ( \partial_\tau  + \tilde{\epsilon}_f  ) z_a \nonumber \\
&+\int_0^\beta d \tau d \tau^\prime \sum_a \Sigma_0(\tau-\tau^\prime)\, 
z_a^{\dag}(\tau) r(\tau)  z_a(\tau^\prime) r(\tau^\prime) \nonumber\\
&- \frac{1}{2} \int d \tau d \tau' D_0(\tau-\tau')\,\frac{2}{N} g^2  T_z(\tau)  T_z(\tau') \nonumber \\
&+\int_0^\beta d\tau \,\sum_a i \Theta \, \,z_a^\dag z_a  + N\int_0^\beta d\tau \,i \Theta \, (r^2/2-  q )
\nonumber \\
& +N\int_0^\beta d\tau \,\lsa(r^2/2- q  ),
\end{align}
where $\tilde{\epsilon}_f = \epsilon_f + \lsa, q=1/N$, and $\Theta = \dot{\theta}+ \Omega_0$.
$\Sigma_0$ is the self-energy due to the interaction with conduction electrons. The imaginary part of $\Sigma_0$ is given by
$- i \Delta_0 \text{sgn}(\epsilon)$, where $\Delta_0$ describes the hybrization amplitude which is usually much larger than the
Kondo energy scale, while
\be
D_0(\tau-\tau') = T \sum_{i\omega} \frac{e^{-i\omega(\tau-\tau')}}{M (\omega^2+\Omega^2)}
\ee
is the bare phonon propagator.
The fermion field $z_a$ can be integrated out in $1/N$ expansion approximation, and  the effective action of bosonic modes
$r, \Theta$ can be obtained.
The parameters $\rsa$ and  $\lsa$ are determined by the saddle point condition, namely that the first order functional derivative
of the effective action of bosonic modes vanishes. The Gaussian fluctuations with respect to this saddle point configuration
 are described by $\delta r = r - \rsa$ and $ i \rsa \Theta $.
These Gaussian fluctuations play important role for  impurity susceptibilities.\cite{coleman}

In analogy with the Kondo spin singlet ground state we look for the orbital pseudospin singlet state,
which is characterized by the condition
\be
\la z_m(\tau) z_n^\dag(\tau') \ra \propto \delta_{mn}.
\ee
The electron spectral function in saddle point approximation is given by
\be
\label{KondoResonance}
A(\epsilon) = \frac{1}{\pi} \frac{1}{(\epsilon-\tilde{\epsilon}_f)^2+\Delta^2},\quad \Delta = \Delta_0 \rsa^2.
\ee
In saddle point approximation the coherent Kondo peak structure is captured, but the incoherent high energy features are missing.
>From Eq. (\ref{KondoResonance}) we can identify $\Delta$ with the (renormalized) Kondo energy scale (thus $\rsa^2 \ll1$)
 and $\tilde{\epsilon}_f$ with the position of Kondo resonance peak. 
$\Delta$ and $\tilde{\epsilon}_f$ are determined by the following saddle point equations (at $T=0$):
\begin{align}
\label{saddle1}
\lsa &= \frac{\Delta_0}{\pi}\ln \left [ \frac{D^2}{\Delta^2 +(\tilde{\epsilon}_f)^2} \right ] \\
&-  \frac{C g^2}{N} \Delta_0 \int^\infty_{-\infty} \frac{d \omega}{2\pi} D_0(i \omega) \frac{\partial \Pi(i \omega)}{\partial \Delta}
\nonumber \\
\label{saddle2}
\frac{\rsa^2}{2} -q &=- \frac{1}{2} + \frac{1}{\pi}\tan^{-1}\left( \frac{\tilde{\epsilon}_f}{\Delta} \right )
- \frac{g^2 C}{2 N}  \frac{ \partial \Pi(i\omega)}{\partial \tilde{\epsilon}_f}.
\end{align}
$D$ is the energy cutoff of the order of bandwidth of conduction electron, and $C$ is a numerical constant of order one. 
$\Pi(i\omega)$ is the polarization function
\be
\label{polarization}
\Pi(i\omega)=-\frac{\Delta}{\pi|\omega| (|\omega|+2 \Delta)}\,
\ln \Big[1+ \frac{|\omega|(2  \Delta  +|\omega|)}{\Delta^2 + (\tilde{\epsilon}_f)^2} \Big].
\ee
If the electron-phonon coupling $g$ is put to zero, Eq.~(\ref{saddle1}) and Eq.~(\ref{saddle2}) reduce to the 
mean-field equation for the Kondo singlet ground state obtained by Coleman. [see Eq. 2.34 of Ref. \onlinecite{coleman}]
The saddle point equations can be solved numerically, but here we are content with the approximate solution for the special
case $q=1/2$ ($N=2$).
In this case Eq.~(\ref{saddle2}) implies that
$\vert \tilde{\epsilon}_f  \vert \ll \Delta$. This also implies $\epsilon_f \sim -\lsa < 0$.
Now Eq.~(\ref{saddle1}) can be approximated to $$
\lsa \approx\frac{2 \Delta_0}{\pi} \ln \frac{D}{\Delta}
-  \frac{C g^2 \Delta_0}{2}  \left [\int_{-\infty}^\infty \frac{d \omega}{2\pi}
 D_0(i \omega) \frac{\partial \Pi(i \omega)}{\partial \Delta} \right ].$$
The effect of the interaction with JT phonon is encapsulated in the second term of the above equation. Using the explicit form of
$\Pi(i\omega)$ we obtain ($\eta$ is a numerical constant of order one)
\be
\label{kondo}
\begin{split}
\Delta &\sim  T_K^{(0)} \exp \left[ - \eta \frac{E_L}{ \Omega} \right ] \quad \text{for}\quad \Omega \gg \Delta, \\
\Delta &\sim  T_K^{(0)} \exp \left[ - \eta \frac{E_L}{ \Omega} \Big( \frac{\Omega}{T_K^{(0)}} \Big )^2 \right ] 
\quad \text{for}\quad \Omega \ll \Delta,
\end{split}
\ee
where
\be
 T_K^{(0)} \sim D \exp \Big[ \frac{\pi}{2} \frac{ \epsilon_f}{\Delta_0} \Big]
\ee
is the unrenormalized Kondo temperature of the infinite-$U$ asymmetric Anderson model obtained 
by Haldane using poor man's scaling.\cite{haldane}
$E_L$ is the lattice relaxation energy defined by
\be
E_L = \frac{g^2}{M \Omega^2}.
\ee
This energy scale is also often called the polaron energy.
We find that the JT electron-phonon interaction \textit{decreases} the Kondo temperature,  which agrees with the previous expectation.
For AH model the Kondo temperature \textit{increases} by the interaction with Holstein phonon.
The Kondo temperature of  the infinite-$U$ AH model obtained with the slave boson method is\cite{hyun2004}
$$T_K \sim T_K^{(0)} \left( 1+ \pi \frac{E_L T_K^{(0)}}{\Delta_0^2} \right),$$
To refine our discussion let us specify the reasonable relative energy scales:
$ \vert \epsilon_f \vert > \Delta_0 > \Omega > ( \Delta \gtrsim E_L)$.
In this parameter regime the renormalization of Kondo temperature in AJT model is much larger than that in AH model.
Apart from the magnitude of renormalization the dependence on various parameters is also markedly different between two models, 
reflecting the essentially different renormalization mechanism.

It is interesting to attempt to understand the decrease of Kondo temperature by JT interaction from alternative approach.
Let us consider the \textit{antiadiabatic limit} where the phonon frequency is the highest energy scale, which 
in our situation implies that $\Omega > \vert \epsilon_f \vert > \Delta_0 > ( \Delta \gtrsim E_L)$.
Clearly this limit does not have much physical relevance, but it does provide some mathematical simplifications.
In the antiadiabatic limit, the bare phonon propagator  becomes local in time $
D_0(\tau -\tau') \to \delta(\tau-\tau')/ M\Omega^2 $.
Now   the integrated JT electron-phonon interaction also becomes local in time
 (note that $f,f^\dag$ are Grassman numbers and are \textit{not}
 operators, in Lagrangian formulation)
$$-E_L T_z T_z  \to  +2 E_L f^\dag_\uparrow f_\uparrow   f^\dag_\downarrow f_\downarrow.$$
 The right-hand side of the above equation is of the same form and of the same \textit{sign} as the local Coulomb repulsion, so that
 it \textit{increases} the local Coulomb repulsion. To address the renormalization by JT interaction in this setup we had better
 consider \textit{finite}-$U$ Anderson model. For simplicity let us take the \textit{symmetric} Anderson model.
>From the Kondo temperature of  the \text{symmetric} Anderson model \cite{haldane} with \textit{renormalized} $U$ one can obtain
\be
T_K  \sim T_K^{(0)} \left( 1 - 2 \frac{E_L}{\Delta_0} \right ),
\ee
which is similar to our result Eq.~(\ref{kondo}) with the replacement of $\Delta_0 \to \Omega$.
Even if the antiadiabatic limit is not physical,  this result gives a supporting evidence for the validity of
our result.

The impurity susceptibilities can be computed by inserting source fields to the 
effective action Eq. (\ref{action:rn1}) and by integrating out $z_a$ and all bosonic modes.\cite{coleman}
The orbital susceptibility is defined by
\be
\chi_{\mathsf{o}}(\tau-\tau') = \la T_z(\tau) T_z(\tau') \ra.
\ee
In the Gaussian approximation one can obtain ($C$ is a numerical constant)
\be
\label{orbitalsusc}
\chi_{\mathsf{o}}(i\omega )=N \Big[ - C \Pi(i\omega) +C^2 g^2 D_0(i\omega) \Pi^2(i\omega) \Big ],
\ee
where $\Pi(i\omega)$ is given in Eq. (\ref{polarization}).
Even though the details are not presented here,  it should be  be noted that
 the  fluctuations of bosonic modes $(\delta r, \Theta)$ do not contribute to the orbital susceptibility.
Evidently, Eq.~(\ref{orbitalsusc}) are the first two terms of random phase approxiamtion (RPA)-type  expansion, 
so that for small $g^2$ we 
can write
\be
\chi_{\mathsf{o}}(i\omega ) \sim \frac{N}{[-C \Pi(i\omega)]^{-1} - g^2 D_0(i\omega)}.
\ee
At high frequency $\omega \gg \Omega$ the renormalization by JT phonon becomes negligible.
In the low frequency limit  $ \omega \to 0$
\be
\label{dcorbital}
\chi_{\mathsf{o}}(i\omega \to 0) \sim \frac{N}{ \pi \Delta - E_L}.
\ee
The result Eq.~(\ref{dcorbital}) clearly demonstrates the JT interaction \textit{strongly enhances} the orbital susceptibility.
This again reflects the fact that the JT phonons couples to the orbital pseudospin channel directly.
In sharp contrast to this result, the \textit{spin} susceptibility of AH model is \textit{not} 
renormalized by the interaction with Holstein phonon  to the same  order of approximation.\cite{hyun2004}

The charge susceptibility is defined by
\be
\chi_{\mathsf{c}}(\tau-\tau') = \la \delta n_f(\tau) \delta n_f(\tau') \ra,
\ee
where
$n_f = \sum_a s_a^\dag s_a = \sum_a z_a^\dag z_a$ and $
 \delta n_f = n_f - \la n_f \ra$.
The fluctuations of bosonic modes play very important role for the charge susceptibility. In the
Gaussian approximation we find that 
in the limit $\omega \to 0$, for the AJT model,
\be
\label{dcharge}
\chi_{\mathsf{c}} (i\omega \to 0) \sim \frac{\pi \Delta}{\Delta_0^2},
\ee
Apparently the JT interaction seems to have no effect on the charge susceptibility, but this is misleading since the
renormalization by JT interaction is reflected in $\Delta$ [Eq. (\ref{kondo})]. Thus  
Eq.~(\ref{dcharge}) signifies that the JT interaction \textit{suppresses} the charge susceptibility.
On the other hand, the charge susceptibility at zero frequency  of the AH model is given by
$$ \chi_{\mathsf{c}}(i\omega \to 0) \sim \frac{\pi T_K^{(0)}}{\Delta_0^2} \Big(1+
\frac{\pi E_L T_K^{(0)}}{\Delta_0^2} \Big),$$
and we can see that the chare susceptibility
 is \textit{enhanced} but by a very small amount.\cite{hyun2004}

In the framework of DMFT there exists a systematic procedure linking the local impurity susceptibility to the
susceptibility of the corresponding lattice system.\cite{dmft1,dmft2} In view of the fact that 
the orbital excitations can be detected by inelastic x-ray scattering, our result on the  orbital
susceptibility  can have important implications for the inelastic x-ray scattering studies of orbital physics.
The renormalization of Kondo temperature can be also experimentally checked, for example, by
the isotope effects experiments on the transition-metal oxide systems and multiferroics.

H.C.L  was  supported by grant No. R01-2005-000-10352-0 from the 
Basic Research Program of the Korea Science and Engineering Foundation.
A part of this work was completed during a stay at the Korea Institute of Advanced Studies in 2005.


\begin{thebibliography}{}
\label{sec:references}
\bibitem{orbital1} Y. Tokura and N. Nagaosa,  Science \textbf{288}, 462 (2000).
\bibitem{orbital2} Y. Tokura, Physics Today \textbf{56} (7), 50 (2003).
\bibitem{park2002} W. Liang, M. Shores, M. Bockrath, J. Long, and H. Park, Nature \textbf{417}, 725 (2002).
\bibitem{clough} D. P. Clougherty,  Phys. Rev. Lett. \textbf{90}, 035507 (2003).
\bibitem{JT} H. A. Jahn and E. Teller, Proc. R. Soc. Lond Ser. A \textbf{161}, 220 (1937);
J. Kanamori, J. Appl. Phys. \textbf{31} (Suppl.), 145 (1960).
\bibitem{meanfield} R. Maezono, S. Ishihara, and N. Nagaosa, Phys. Rev. B \textbf{58}, 11583 (1998).
\bibitem{ME}  A. B. Migdal, Sov. Phys. JETP \textbf{7}, 996 (1958); G. M.
Eliashberg, \textit{ibid.} \textbf{11}, 696 (1960).
\bibitem{dmft1} A. Georges, G. Kotliar, W. Krauth, and M. J. Rosenberg,
 Rev. Mod. Phys. \textbf{68}, 13 (1996).
\bibitem{dmft2} G. Kotliar and D. Vollhardt, Physics Today \textbf{57} (3), 53 (2004).
\bibitem{urushibara} A. Urushibara, Y. Moritomo, T. Arima, A. Asamitsu, G. Kido, and Y. Tokura,  Phys. Rev. B \textbf{51}, 14103 (1995).
\bibitem{holstein} T. Holstein, Ann. Phys. \textbf{8}, 325  (1959).
 \bibitem{hewson} A. C. Hewson and D. Meyer, J. Phys.:Condens. Matter \textbf{14},
 427 (2002).
\bibitem{hewson2} D. Meyer, A. C. Hewson, and R. Bulla, Phys. Rev. Lett. 
\textbf{89}, 196401 (2002).
\bibitem{jpc} G. S. Jeon, T. H. Park, and H. Y. Choi, Phys. Rev. B \textbf{68}, 045106 (2003).
\bibitem{ours} H. C. Lee and H. Y.  Choi, Phys. Rev. B \textbf{69}, 075109 (2004).
\bibitem{hyun2004} H. C. Lee and H. Y. Choi, Phys. Rev. B \textbf{70}, 085114 (2004).
\bibitem{thebook} A. C. Hewson, \textit{ The Kondo Problem to Heavy Fermions} (Cambridge University Press, Cambridge, 1993).
\bibitem{coleman} P. Coleman, Phys. Rev. B \textbf{35}, 5072 (1987).
\bibitem{read} N. Read and D. M. Newns, J. Phys. C. \textbf{16}, 3273 (1983).
\bibitem{read2} N. Read and D. M. Newns, J. Phys. C. \textbf{29}, L1055 (1983).
\bibitem{SG84} K. Sch\"onhammer and O. Gunnarsson, Phys. Rev. B \textbf{30}, 3141 (1984).
\bibitem{haldane} F. D. M. Haldane, Phys. Rev. Lett. \textbf{40}, 416 (1978).
\bibitem{maekawa} S. Ishihara and S. Maekawa, Phys. Rev. B \textbf{62}, 2338 (2000).
\end{thebibliography}
\end{document}